\documentclass[12pt,a4paper,final]{iopart}

\usepackage{float}
\usepackage{iopams}
\usepackage{graphicx}
\usepackage{cite}
\usepackage{xcolor}
\usepackage[breaklinks=true,colorlinks=true,linkcolor=blue,urlcolor=blue,citecolor=blue]{hyperref}

\begin{document}

\title{Electron heating mode transitions in radio-frequency  driven micro atmospheric pressure plasma jets in He/O$_{2}$: A fluid dynamics approach}

\author{Yue Liu$^{1,3}$, Ihor Korolov$^{1}$, Torben Hemke$^{1}$, Lena Bischoff$^{1}$, Gerrit Hübner$^{1}$, Julian Schulze$^{1,2}$, and Thomas Mussenbrock$^{1}$}
\address{$^{1}$ Department of Electrical Engineering and Information Science, Ruhr-University Bochum, D-44780, Bochum, Germany\\
$^{2}$ Key Laboratory of Materials Modification by Laser, Ion and Electron Beams, School of Physics, Dalian University of Technology, Dalian, 116024, China\\
$^{3}$ Author to whom correspondence should be addressed}

\medskip

\eads{liu@aept.ruhr-uni-bochum.de}

\begin{abstract}
A two-dimensional fluid model is used to investigate the electron heating dynamics and the production of neutral species in a capacitively coupled radio-frequency micro atmospheric pressure helium plasma jet -- specifically the COST jet -- with a small oxygen admixture. Electron heating mode transitions are found to be induced by varying the driving voltage amplitude and the O$_2$ concentration numerically and experimentally. The helium metastable density, and the charged species densities are highly relevant to the electron heating dynamics. By analyzing the creation and destruction mechanisms of the negative ions, we find that the generation of negative ions strongly depends on the O$_2$ concentration. The increase of the electronegativity with the increasing O$_2$ concentration leads to an enhancement of the bulk drift electric field. The distributions of the different neutral species densities along the direction of the gas flow inside the jet, as well as in the effluent differ a lot due to the relevant chemical reaction rates and the effect of the gas flow. The simulated results show that a fluid model can be an effective tool for qualitative investigations of micro atmospheric pressure plasma jets.
\end{abstract}

\vspace{2pc}

\section{Introduction}
Radio-frequency (RF) micro atmospheric pressure plasma jets ($\mu$-APPJs) have become an attractive plasma source for surface treatment \cite{laroussiLowTemperaturePlasmaBased2005a,beckerMicroplasmasApplications2006a,adamovich2017PlasmaRoadmap2017a,penkovReviewRecentApplications2015,babayanDepositionSiliconDioxide1998a,ichikiLocalizedUltrahighrateEtching2003a} and, in particular, for biomedical applications \cite{kimCharacterizationAtmosphericPressure2009,gravesLowTemperaturePlasma2014a,weltmannPlasmaMedicineCurrent2016,kongPlasmaMedicineIntroductory2009,bekeschusPlasmaJetKINPen2016}. Such jets are usually operated in helium or argon with a small admixture of molecular gases, such as nitrogen, oxygen or combinations of both. The control of the production of reactive oxygen and nitrogen species (RONS) within the discharge volume is crucial for these applications. One method is the use of voltage waveform tailoring \cite{heilPossibilityMakingGeometrically2008a,schulzeElectricalAsymmetryEffect2011a}, which has been shown the capability of tuning the dynamics of electron energy distributions (EED) \cite{gibsonDisruptingSpatiotemporalSymmetry2019,bischoffExperimentalComputationalInvestigations2018b,korolovControlElectronDynamics2019,korolovHeliumMetastableSpecies2020,liuMicroAtmosphericPressure2020}. Typically, the EED deviates from a Maxwellian distribution particularly for high energies even at atmospheric pressure since the energy relaxation length can be larger than length scale of the reduced electric field gradients. Furthermore, the energy relaxation frequency can be faster than the temporal change of the reduced electric field.

Initial investigations on electron power absorption dynamics in $\mu$-APPJs were performed in nominally pure argon or helium. The electron heating modes were assumed to be similar to those of low pressure capacitive discharges, i.e., the $\alpha$-mode and the $\gamma$-mode \cite{izaElectronKineticsRadioFrequency2007a,niemiRoleHeliumMetastable2011,niermannSpatialDynamicsHelium2011b,dunnbierStabilityExcitationDynamics2015}. The $\alpha$ to $\gamma$ transition was later discussed in the context of adding impurities \cite{chirokovNumericalExperimentalInvestigation2009,kawamuraParticleincellGlobalSimulations2014a}. Hemke \textit{et al.} \cite{hemkeIonizationBulkHeating2012a} firstly pointed out based on results of Particle-in-Cell/Monte Carlo Collision (PIC/MCC) simulations for a pure helium discharge that the ionization dynamics is mainly produced by Ohmic heating in the so-called $\Omega$-mode. It was demonstrated that even a small impurity in the noble gases can change the dominant ionization path to Penning ionization, leading to a decrease of the breakdown voltage \cite{martensDominantRoleImpurities2008,raduFrequencyVoltageDependence2003}. Experimentally, phase resolved optical emission spectroscopy (PROES) was used to study the dynamics of energetic electrons based on the wavelength integrated optical emission \cite{gathenSpatiallyResolvedDiagnostics2008,benediktPhaseResolvedOptical2010,reuterDetectionOzoneMHz2012,schaperElectronDynamicsRadioFrequencyDriven2011} and the use of selected emission lines (Ar: 750 nm \cite{dunnbierStabilityExcitationDynamics2015}, O: 844 nm \cite{grebEnergyResolvedActinometry2014,waskoenigAtomicOxygenFormation2010b}). Bischoff \textit{et al.} \cite{bischoffExperimentalComputationalInvestigations2018b} proposed that the 706.5 nm helium line can be used to probe the ionization dynamics in helium when using a small nitrogen admixture. Electron power absorption mode transitions were observed as well in that work by both experiments and PIC/MCC simulations for $\mu$-APPJs operated in He/N$_{2}$ mixtures.

Low pressure electropositive radio frequency capacitively coupled plasmas (RF-CCPs) can operate in two modes, the $\alpha$-mode and the $\gamma$-mode \cite{belenguerTransitionDifferentRegimes1990}. Similarly, RF driven $\mu$-APPJs can also operate in two modes, the aforementioned $\Omega$-mode and the Penning-mode. In the $\Omega$-mode, electrons are accelerated by a bulk electric field while having a high neutral collision frequency leading to a decreased electron conductivity in spite of the electron conduction current being high. Although the spatio-temporal ionization dynamics of the $\Omega$-mode is similar to those of the $\alpha$-mode, the physical mechanism is different. In the $\alpha$-mode in low pressure RF-CCPs, energetic electrons are mainly generated by sheath expansion. The electric field adjacent to the sheath can be either the ambipolar field or a drift field, or even a combination of both due to low electron density\cite{brinkmannElectricFieldCapacitively2015,brinkmannElectronHeatingCapacitively2015,schulzeSpatiotemporalAnalysisElectron2018}.

There is also a fundamental difference between the Penning-mode in $\mu$-APPJs and the $\gamma$-mode in low pressure RF-CCPs. The largest electron impact ionization rate occurs inside the sheath in both modes at maximum sheath extension. In the $\gamma$-mode this ionization is produced by secondary electron emission from the electrodes. Whereas in the $\mu$-APPJs, the maximum ionization in the Penning-mode is caused by energetic electrons producing highly excited states of the rare gas, followed by Penning ionization, for example, in He/O$_{2}$ mixtures, He* + O$_{2}$ $\rightarrow$ e + He + O$_{2}^{+}$.

Strongly electronegative capacitive RF discharges at low pressure, such as O$_2$ or CF$_4$, operate in a drift-ambipolar mode \cite{schulzeIonizationDriftAmbipolar2011b}, dominated by a drift electric field in the bulk and an ambipolar electric field near the sheath edge. The drift field here results from a low electron density in the bulk due to depletion by attachment rather than the high collision frequency in the $\Omega$-mode. The ambipolar field is a consequence of a local maximum in the electron density at the sheath edge.

Another important aspect of $\mu$-APPJs for any kind of applications is the production of reactive species\cite{hemkeSpatiallyResolvedSimulation2011b}, which are highly relevant for chemical reactions. Complex plasma chemistries, such as He/O$_2$ mixtures, consist of hundreds of reactions. A sensitivity analysis \cite{turnerUncertaintySensitivityAnalysis2015}, as well as a discussion of uncertainties and errors \cite{turnerUncertaintyErrorComplex2015a} induced by individual reactions were performed in the frame of a global model. Simulated species densities (atomic oxygen densities \cite{waskoenigAtomicOxygenFormation2010b}, ozone densities \cite{wijaikhumAbsoluteOzoneDensities2017a}) were benchmarked against experiments. The major reactions that lead to the generation and loss of these two species were discussed.  Reactive species generations were also found to be affected by discharge parameters. The pressure dependence of O$_2$(a$^{1}\Delta_{g}$) densities \cite{babaevaProductionO2D12006a}, the effect of power pulse control on O$_2$(a$^{1}\Delta_{g}$) production \cite{babaevaO2D1Production2007b}, as well as the influence of the electrode configuration on RONS production \cite{lietzElectrodeConfigurationsAtmospheric2018} were investigated based on two-dimensional fluid simulations. Additional reactive neutral species are generated through interactions between the effluent and its surroundings. Parametric studies on RONS production in a APPJ flowing into humid air \cite{norbergFormationReactiveOxygen2015}, or interacting with water \cite{norbergAtmosphericPressurePlasma2018} were performed based on a two-dimensional plasma hydrodynamics model.

In this work, we investigate a radio frequency capacitively coupled micro atmospheric pressure He/O$_2$ plasma jet based on a two-dimensional fluid dynamics approach. The purpose of this study is to address the effect of voltage amplitude and molecular reactive admixtures on the electron heating mode transition, the charged species dynamics and the neutral species densities. Based on the simulation results and comparisons to experimental results, it is shown that a fluid model can qualitatively describe the variation tendency induced by control parameters and capture the main physical mechanism in such micro atmospheric pressure discharges. The remaining content of this paper is structured as follows: a brief introduction of the simulation model is provided in section \ref{section2}, followed by discussions of results in section \ref{section3}. Finally, conclusions are drawn and prospects are made.  

\section{\label{section2}Model}
\subsection{\label{section2.1}Description of the simulation model}
The simulations in this investigation were performed using the computational modeling platform \textit{nonPDPSIM} developed by Mark Kushner \cite{kushnerModelingMicrodischargeDevices2004a,kushnerModellingMicrodischargeDevices2005a}. It is a two-dimensional multi-species fluid dynamics code based on an unstructured grid, used for medium to high pressure weakly ionized plasmas. The model and prior applications have been previously discussed in detail \cite{kushnerModelingMicrodischargeDevices2004a,kushnerModellingMicrodischargeDevices2005a}. \textit{nonPDPSIM} has been successfully applied to the study of micro atmospheric pressure plasma jets \cite{babaevaProductionO2D12006a,babaevaO2D1Production2007b,hemkeSpatiallyResolvedSimulation2011b,niermannSpatialDynamicsHelium2011b,norbergFormationReactiveOxygen2015,lietzElectrodeConfigurationsAtmospheric2018,norbergAtmosphericPressurePlasma2018}. Here, only a brief description of \textit{nonPDPSIM} is provided.

For each charged species, the particle conservation equation is solved. The particle flux is expressed in terms of the drift-diffusion approximation. In order to allow for the non-Maxwellian characteristics of the electron energy distribution function (EEDF), the electron transport coefficients, as well as the rate constants in the source and loss terms, are obtained by solving a 0-dimensional Boltzmann equation based on the 2-term approximation. These generated coefficients are firstly tabulated as a function of the reduced electric field, subsequently altered in dependence of the mean electron energy, or the effective mean electron temperature. The effective mean electron temperature is obtained by solving the electron energy conservation equation. The fluid equations of charged species are coupled to incompressible (or if needed compressible) Navier-Stokes equations, which are used to describe the neutral species transport. Poisson’s equation is solved for the electric potential.

\begin{figure}[t]
    \centering
    \includegraphics{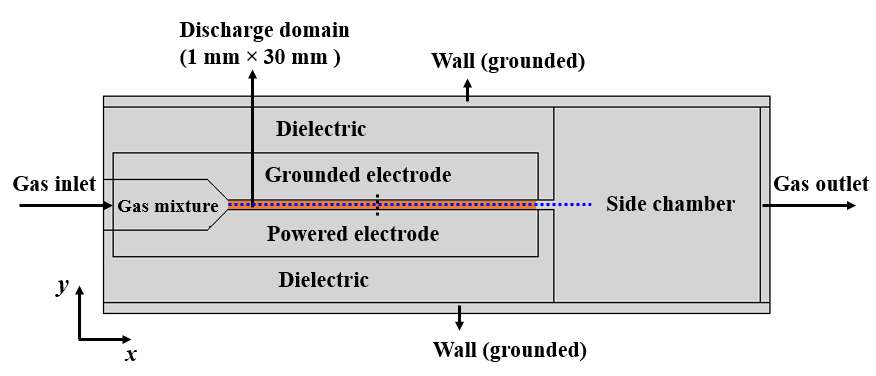}
    \caption{Schematic of the simulation geometry based on the COST reference micro-plasma jet introduced in \cite{goldaConceptsCharacteristicsCOST2016}. The black dotted line perpendicular to the electrodes represents positions where results shown in figures \ref{fig2}-\ref{fig8} are taken. The blue dotted line parallel to the electrodes represents positions where results shown in figure \ref{fig9} and \ref{fig10} are taken.}
    \label{fig1}
\end{figure}

A schematic of our simulation domain is shown in figure \ref{fig1}. It is based on the COST reference micro-plasma jet introduced in \cite{goldaConceptsCharacteristicsCOST2016}. In our simulations, a finite volume of the model is considered, whose edges are grounded. The powered electrode is at the bottom and the grounded electrode is at the top. Two dielectrics are placed between the respective electrode and the adjacent grounded wall, with a relative permittivity equal to 4. Gases flow in on the left side and are mixed in the gas mixing volume. It is 5 mm in height. Therefore, the discharge cannot be ignited in this region at voltages typically used to drive the jet. After the mixing region there is the discharge channel of 30 mm in length and 1 mm in height. A side chamber is located after the discharge domain for the effluent before gases flow out on the right. The unstructured mesh of design includes approximately 12000 nodes for the plasma region.

The discharge is operated in He/O$_2$ mixtures at atmospheric pressure. The gas flow rate can significantly affect plasma discharges and species generations. A fast gas flow will carry most of the generated neutral species outside the jet, meanwhile, it will cause turbulence in the jet. Those together lead to unsteady discharges. While a slow gas flow will lead to reactive species fully built up in the jet so that the treatment distance is limited. As a result, the gas flow is influenced by the length of the jet, and it is required to generate high fluxes of reactive species at the nozzle in an energy efficient way. Considering the feature size of a COST jet, it usually works in a gas flow rate approximate 1 slm \cite{hemkeSpatiallyResolvedSimulation2011b,goldaConceptsCharacteristicsCOST2016,knakeAbsoluteAtomicOxygen2008a,waskoenigAtomicOxygenFormation2010b,bischoffExperimentalComputationalInvestigations2018b,korolovControlElectronDynamics2019,korolovHeliumMetastableSpecies2020}. The gas flow is fixed at 1 slm in simulations. The oxygen concentration ratio is set to 0.05\% , 0.25\%, and 0.5\% for different simulation cases. The voltage source is fixed at a frequency of 13.56 MHz and the voltage amplitude is varied from 400 V to 600 V. The included species are ground state neutral species He, O$_2$, O, O$_3$, excited state neutral species O$_2$(v=1-4) (first four vibrational levels of O$_2$), O$_3$(v), O$_2$(a$^{1}\Delta_{g}$), O$_2$(b$^{1}\Sigma_{g}^{+}$), O($^1$D), He* (ensemble of He(2$^3$S) and He(2$^1$S)), positive ions O$_{2}^{+}$, O$^{+}$, He$^{+}$, negative ions O$^{-}$, O$_{2}^{-}$, O$_{3}^{-}$, and electrons. For electron impact collisions with helium atoms, cross sections of the elastic \cite{CrossSectionsExtracted}, the excitations \cite{CrossSectionsExtracted,CrossSectionsExtractedc} and the ionization process \cite{CrossSectionsExtracted} are considered. For electron impact collisions with oxygen molecules, the reactions and the corresponding cross sections proposed by Gudmundsson et al. \cite{gudmundssonBenchmarkStudyCapacitively2013} are considered. These cross sections are used to generate transport coefficients by solving the Boltzmann equation. Reactions between electrons and other neutral species are neglected, since those neutral densities are at least one order of magnitude lower than the molecular oxygen density. Reactions between heavy species (ions and neutrals) are based on Turner \cite{turnerUncertaintyErrorComplex2015a}. The number of total reactions is reduced according to the sensitivity analyses \cite{turnerUncertaintySensitivityAnalysis2015}. The chemical reactions considered in this work are the same as those listed in \cite{liuMicroAtmosphericPressure2020}. Surface coefficients for neutral species are treated as a loss probability \cite{staffordO2D1Production2004b,waskoenigAtomicOxygenFormation2010b}. The coefficients are identical to those listed in \cite{liuMicroAtmosphericPressure2020}. The ion induced secondary electron emission coefficients are chosen to be 0.2, 0.06, and 0.1 for He$^{+}$, O$_{2}^{+}$, and O$^{+}$ respectively based on the formula given in \cite{raizerGasDischargePhysics1991}.

\subsection{\label{section2.2}Description of the experimental set-up}
The experimental set-up has been introduced in detail by Bischoff \textit{et al.} \cite{bischoffExperimentalComputationalInvestigations2018b} and Korolov \textit{et al.} \cite{korolovControlElectronDynamics2019}. Here only a brief description is demonstrated. Experiments are performed using a RF driven COST-jet \cite{goldaConceptsCharacteristicsCOST2016} operated in He combined with different O$_2$ admixtures. The jet consists of two parallel stainless-steel electrodes of 30 mm in length. The gap between the electrodes is 1 mm. 5.0 purity helium and oxygen gases are used. The gas flow of He is fixed at 1 slm and the O$_2$ flow is varied from 0.5 sccm to 5 sccm. The RF voltage is applied to the powered electrode by a power generator via a matching network. The voltage waveform at the powered electrode is measured by a voltage probe (Tektronix P6015A with a bandwidth of 75 MHz). Phased resolved optical emission spectroscopy (PROES) is used to observe the helium emission line at 706.5 nm via an interference filter at 700 nm wavelength and 15 nm of full width at half maximum. The threshold of the corresponding electron impact helium excitation reaction is 22.7 eV. In this case only energetic electrons are detected as discussed by Bischoff \textit{et al.} \cite{bischoffExperimentalComputationalInvestigations2018b}. The spatial-temporally resolved emission is recorded by an ICCD camera with a gate width of 1 ns. The measurements are taken at the position of -10 mm from the nozzle. (The coordinate system is shown in figure \ref{fig9}.) The image resolution between the electrode gap corresponds 149 pixels. To monitor the impurity level, time integrated optical emission spectroscopy (OES) is also conducted by a Universal Serial Bus (USB) grating spectrometer.

\section{\label{section3}Results}
\begin{figure}[t]
    \centering
    \includegraphics[width=1.0\textwidth]{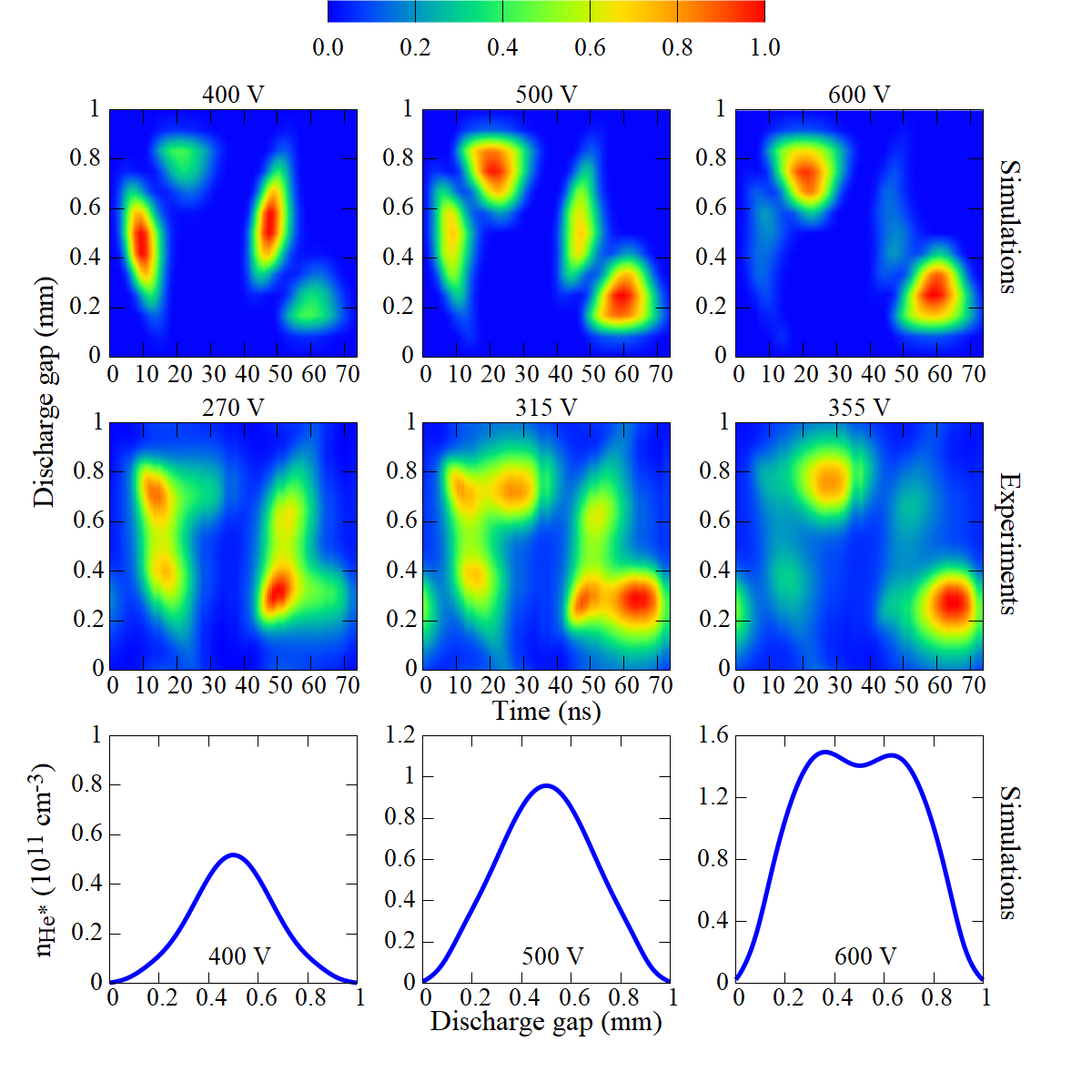}
    \caption{Spatio-temporally resolved He(3$^3$S) excitation rates from simulations (first row) and from experiments (second row), and the computationally obtained time-averaged helium metastable (ensemble of the triplet and the singlet) density profiles (third row) between the two electrodes as a function of the driving voltage amplitude. The gas flow is fixed at 1 slm and the oxygen concentration is kept constant at 0.05\%.}
    \label{fig2}
\end{figure}

The first row of figure \ref{fig2} shows the computational spatio-temporally resolved He(3$^3$S) excitation rates as a function of the driving voltage amplitude at 400 V, 500 V, and 600 V (different columns), with the gas flow fixed at 1 slm and the oxygen concentration kept constant at 0.05\%. The results are taken at the center of the discharge channel (marked by the black dotted line in figure \ref{fig1}). The maximal excitation rates are different for each case. The results are normalized by the respective maximum. As the driving voltage amplitude increases, the spatio-temporal dynamics of the excitation rate changes, i.e., an electron heating mode transition is induced. At lower voltage, the majority of energetic electrons are generated in the plasma bulk when the sheaths are expanding; while at higher voltages, strong electron impact excitation rates appear inside the sheaths at the time of maximal sheath voltage. It should be noted that even though the electric field in the sheath is much stronger than in the bulk at 400 V, the electron density in the bulk is much higher compared to the sheath. As a consequence, most energetic electrons (above 22.7 eV) are generated by the acceleration due to the bulk drift field and the discharge is operated in the $\Omega$-mode \cite{hemkeIonizationBulkHeating2012a,bischoffExperimentalComputationalInvestigations2018b}. 

When the voltage is increased to 600 V, the electric field in the sheath becomes strong enough to dominate the production of energetic electrons, causing the discharge to be operated in the Penning-mode \cite{hemkeIonizationBulkHeating2012a,bischoffExperimentalComputationalInvestigations2018b}. As can be seen in figure \ref{fig2}, the strong electron impact excitation rates occur when the sheaths expand to the maximum, i.e. when the electric field in the sheath is strongest. Those electrons mainly originate from Penning ionization inside the sheath. The same electron heating mode transitions are found via PROES measurements in a lower voltage amplitude range from 270 V to 355 V, as shown in the second row of figure \ref{fig2}. Each case is normalized by the respective maximum. A larger grounded electrode compared to the powered electrode leads to the weak asymmetry of the patterns in the experimental results. The vertical stripes in experimental results are due to synchronization issues of the camera. As the camera is directly triggered by the applied waveform, at some moments the internal delay generator of the camera is not very stable, thus, leading to an inaccurate set of the delay time. The difference of the working voltage amplitude between simulations and experiments is due to the ignorance of the electron kinetic effects in fluid simulations, particularly for the electrons at high energy, such as the electrons generated from Penning ionization and surface emissions, which are then accelerated by the electric field. On the other hand, the total number of non-uniform unstructured cells is limited to a reasonable number to save the computational cost. Due to the far larger effective electrode length than the electrode gap, a relatively coarse meshing along the electrode gap is used, which can lead to a reduced precision of the spatial resolution between the electrodes. However, the simulation results still show a qualitative agreement with the experiments. It is very important to understand such a mode transition, since it results in a lot of plasma parameter variations. For example, as shown in the third row of figure \ref{fig2}, the time-averaged helium metastable density (simulated) is increased, with two peaks near the sheath edges at 600 V. Helium metastables are generated via electron impact excitations, while mainly destructed by Penning ionizations with O$_2$. Both reactions proceed fast. Since O$_2$ is uniformly distributed, the helium metastable density profile depends on the generation rate distribution, i.e., it is the highest at the center of the discharge gap in the $\Omega$-mode, while two peaks near the sheath edges are formed in the Penning-mode.

\begin{figure}[t]
    \centering
    \includegraphics[width=1.0\textwidth]{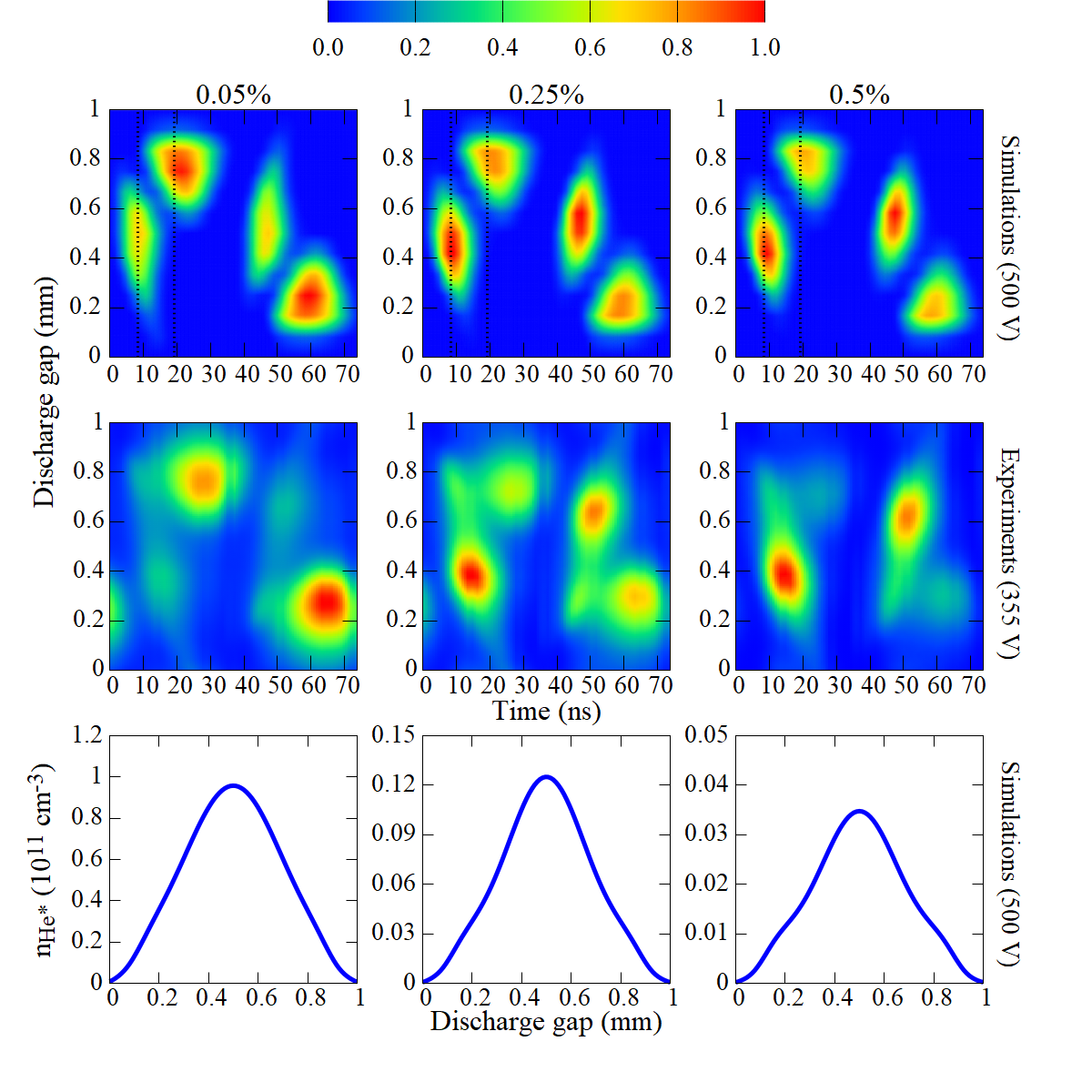}
    \caption{Spatio-temporally resolved He(3$^3$S) excitation rates from simulations (first row) and from experiments (second row), and the computationally obtained time-averaged helium metastable (ensemble of the triplet and the singlet) density profiles (third row) between the two electrodes as a function of the O$_2$ concentration. The gas flow is 1 slm. The applied voltage amplitude is 500 V in simulations and 355 V in experiments. The black dotted lines in the first row correspond to the moment at which results are taken and shown in figure \ref{fig8} for each case.}
    \label{fig3}
\end{figure}

The first row of figure \ref{fig3} shows the computational spatio-temporally resolved He(3$^3$S) excitation rates as a function of the O$_2$ concentration. The voltage amplitude is fixed at 500 V. The results are taken again at the center of the gas flow channel (marked by the black dotted line in figure \ref{fig1}). The maximal excitation rates are different for each case. The results are normalized by the respective maximum. For the 0.05\% O$_2$ concentration case, the excitation rates are stronger in the sheaths when the sheaths expand to the maximum, indicating that the discharge is predominantly operated in the Penning-mode. Increasing the O$_2$ concentration induces the discharge to be predominantly operated in the $\Omega$-mode, since the excitation rates are stronger in the plasma bulk when the sheaths are expanding. Such a transition is also shown by PROES measurements in the second row of figure \ref{fig3} at 355 V. The slight asymmetry is caused by the larger area of the grounded electrode. The reason for the use of higher voltage amplitudes in the simulations has been discussed above. This transition is caused by the combination of two effects: electronegativity and collisions. More electronegative gas can lead to more negative ions and a lower electron density in the bulk (shown below), leading to a stronger bulk drift electric field, since the electron conductivity is inversely proportional to the electron density. This is similar to the drift pattern of the drift-ambipolar mode \cite{schulzeIonizationDriftAmbipolar2011b} in low pressure RF strongly electronegative capacitive discharges. Besides, the destruction rate of He* is enhanced by the increased O$_2$ density. However, the majority of electrons generated via the Penning ionization cannot be accelerated to high energy due to the more frequent inelastic collisions in the presence of more molecular gas, leading to a decreased population of He*. Correspondingly, the helium metastable density decreases significantly by adding more O$_2$ as shown in the third row of figure \ref{fig3}.

\begin{figure}[t]
    \centering
    \includegraphics[width=1.0\textwidth]{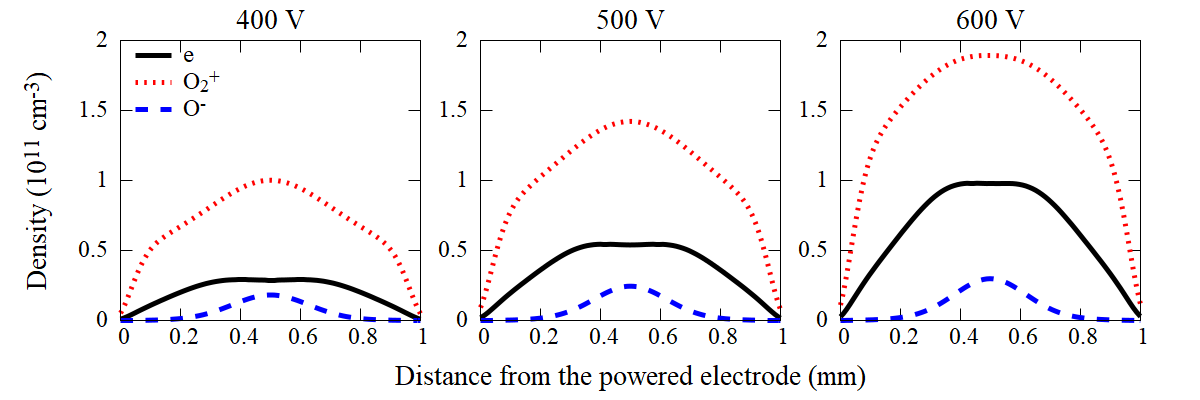}
    \caption{Simulated time-averaged charged species density profiles between the electrodes as a function of the voltage amplitude at the longitudinal position indicated in figure \ref{fig1}. The O$_2$ concentration is kept constant at 0.05\%.}
    \label{fig4}
\end{figure}

\begin{figure}[t]
    \centering
    \includegraphics[width=0.5\textwidth]{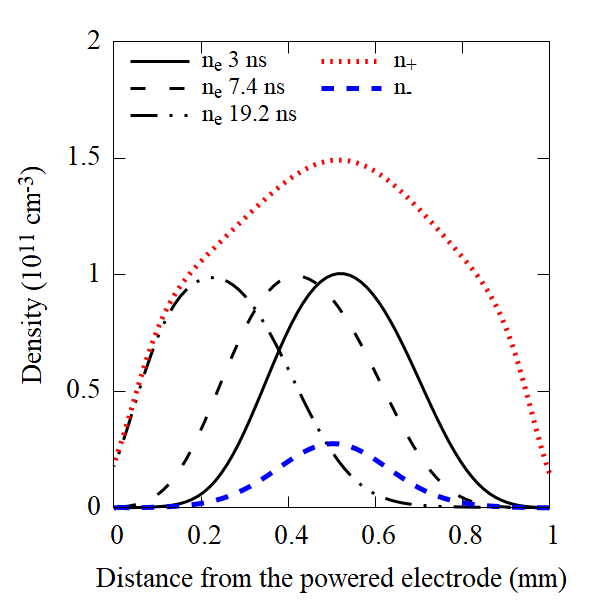}
    \caption{Computational electron, positive ion and negative ion densities at different moments within one RF period at the longitudinal position indicated in figure \ref{fig1}. The voltage amplitude is 500 V and the O$_2$ concentration is kept constant at 0.05\%.}
    \label{fig5}
\end{figure}

As shown above, the helium metastable density is highly relevant to the electron heating dynamics, and so are the charged species densities. Figure \ref{fig4} shows the simulated time-averaged charged species density profiles between the electrodes as a function of the voltage amplitude. The O$_2$ concentration is kept constant at 0.05\%. All the results are taken at the center of the discharge channel (marked by the black dotted line in figure \ref{fig1}). In such cases, merely electrons, O$_{2}^{+}$ and O$^{-}$ are dominant, while the other charged species are negligible. Their densities are increased by increasing the voltage amplitude, but the variation for O$^{-}$ is weak. This is due to the competition between the enhanced major generation (electron impact dissociation attachment) and destruction (recombination with O$_{2}^{+}$ and reactions with oxygen neutrals). It can be seen that the time-averaged quasi-neutrality is broken, since the O$_{2}^{+}$ density is higher than the sum of electron and O$^{-}$ densities, for example, at the center of the discharge gap. We believe that it is an averaging effect. The boundary loss for electrons is pronounced due to a small discharge gap (1 mm) and the strong oscillation driven by the RF electric field, which results in a narrow profile of the electron density at each moment. Ion densities are almost time independent in steady state, thus, the spatio-temporal profiles of ion densities are not shown. The instantaneous quasi-neutrality is fulfilled locally (shown in figure \ref{fig5}), while the mean electron density is decreased by averaging over one RF period.

\begin{figure}[t]
    \centering
    \includegraphics[width=1.0\textwidth]{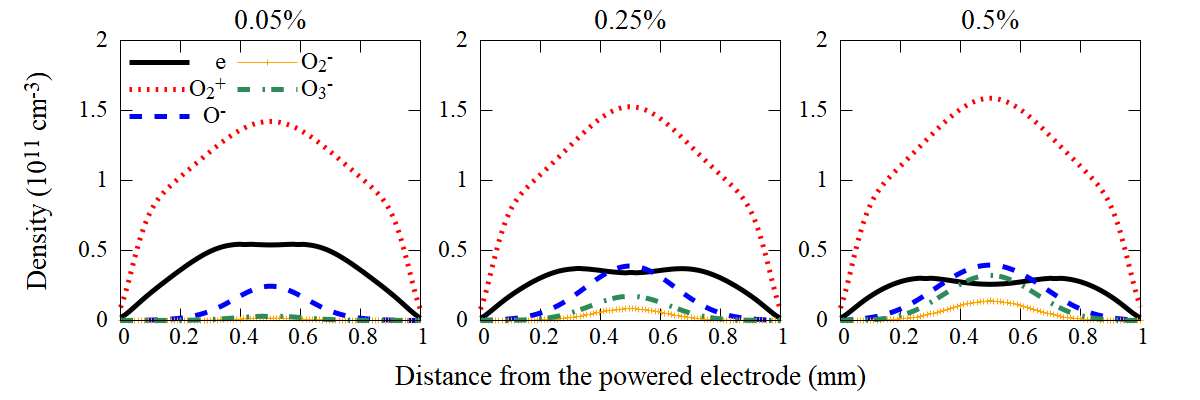}
    \caption{Simulated time-averaged charged species density profiles between the electrodes as a function of the O$_2$ concentration at the longitudinal position indicated in figure \ref{fig1}. The voltage amplitude is kept at 500 V.}
    \label{fig6}
\end{figure}

\begin{figure}[t]
    \centering
    \includegraphics[width=0.5\textwidth]{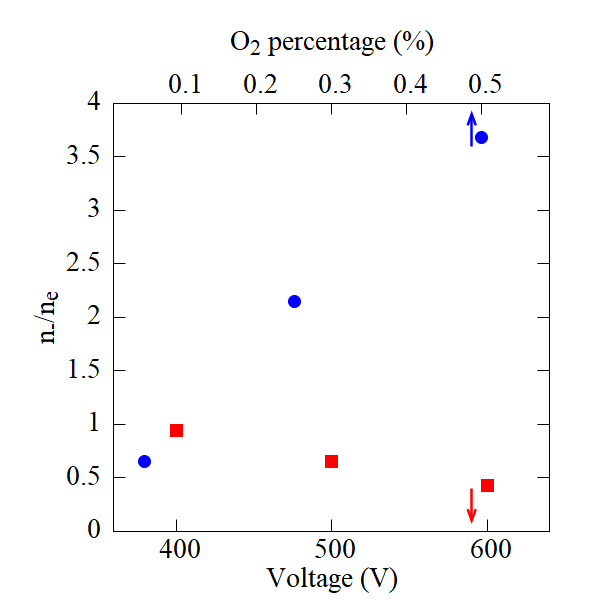}
    \caption{Ratios between the sum of the time-averaged negative ion densities and the time-averaged electron densities at the center of the discharge gap as a function of the voltage amplitude (red square, 0.05\% O$_2$ concentration) and the O$_2$ concentration (blue dot, the voltage amplitude of 500 V). The densities are taken at the longitudinal position indicated in figure \ref{fig1}. }
    \label{fig7}
\end{figure}

Figure \ref{fig6} shows the simulated time-averaged charged species density profiles between the electrodes as a function of the O$_2$ concentration. The voltage amplitude is kept constant at 500 V. All the density profiles are shown at the center of the discharge channel (marked by the black dotted line in figure \ref{fig1}). At low O$_2$ concentration (0.05\%), the electron density is higher than the negative ion densities. The major negative ion is O$^{-}$, while O$_{2}^{-}$ and O$_{3}^{-}$ are negligible. As the O$_2$ flow is increased, the electron density decreases, and the hump in the density profile becomes significant due to the enhancement of the negative ion population at the center of the electrode gap. At 0.5\% O$_2$ concentration, the O$^{-}$ and O$_{3}^{-}$ densities predominate over the electron density. The O$^{-}$ density increases slightly compared to the significant enhancement of the O$_{2}^{-}$ and O$_{3}^{-}$ densities. To understand such a variation of the negative ion density as a function of the O$_2$ concentration, it is necessary to investigate the major chemical reactions leading to the construction and destruction of those negative ions. We estimate the production rate and loss rate induced by each relevant reaction based on the time-averaged density of each species and the corresponding rate constant. By comparing those estimated production and loss rates, we find that O$^{-}$, O$_{2}^{-}$ and O$_{3}^{-}$ are mainly produced by the reactions

\begin{equation}
\mathrm{e} + \mathrm{O}_{2} \rightarrow \mathrm{O} + \mathrm{O}^{-},
\label{eq1}
\end{equation}

\begin{equation}
\mathrm{e} + \mathrm{O}_{2} + \mathrm{He} \rightarrow \mathrm{O}_{2}^{-} + \mathrm{He},
\label{eq2}
\end{equation}

\begin{equation}
\mathrm{O}^{-} + \mathrm{O}_2 + \mathrm{He} \rightarrow \mathrm{O}_{3}^{-} + \mathrm{He},
\label{eq3}
\end{equation}
while the destruction rates, as mentioned above for negative ions, differ merely slightly between those different species. It can be seen that the generation of negative ions is proportional to the O$_2$ concentration. However, reaction (\ref{eq3}) also corresponds to a destruction mechanism of O$^{-}$, causing the O$^{-}$ density to increase only slightly by increasing the O$_2$ concentration. Such variations of the negative ion and electron densities as a function of the voltage amplitude and the O$_2$ concentration can affect the electronegativity. Here the electronegativity is defined as the ratio between the sum of the time-averaged negative ion densities and the time-averaged electron density at the center of the discharge gap. As shown in figure \ref{fig7}, this ratio decreases as the voltage amplitude increases (red square). However, a significant increase of the electronegativity results from the increasing O$_2$ concentration (blue dot).

\begin{figure}[t]
    \centering
    \includegraphics[width=0.8\textwidth]{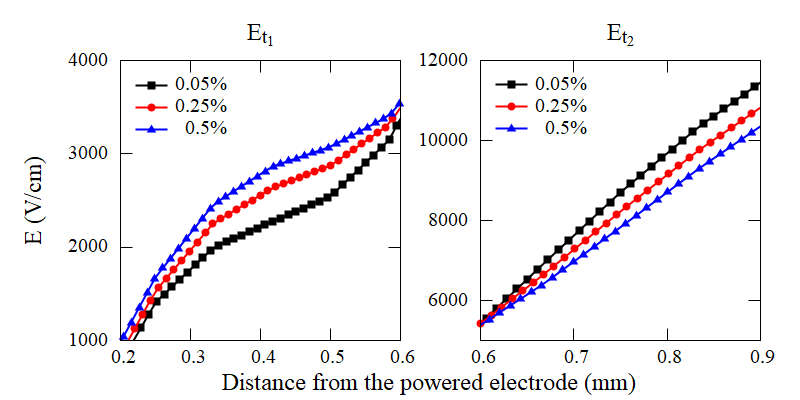}
    \caption{Simulated electric field perpendicular to the electrode along the electrode gap at t$_{1}$ = 8.3 ns when the excitation rate is maximal in the bulk (left) and t$_{2}$ = 19.2 ns when the excitation rate is maximal in the sheath (right) as a function of the O$_2$ concentration at the fixed voltage amplitude of 500 V. The corresponding moments are marked by black dotted lines in the first row of figure \ref{fig3}. The results are taken along the black dotted line indicated in figure \ref{fig1}.}
    \label{fig8}
\end{figure}

As a consequence of the variation of the electronegativity, the electric field perpendicular to the electrode changes with the O$_2$ concentration. In figure \ref{fig8}, the left figure shows the electric field component perpendicular to the electrodes at t$_1$ = 8.3 ns corresponding to the moment of the maximal excitation rate in the bulk (marked in the first row of figure \ref{fig3}). We restrict the region from 0.2 mm to 0.6 mm to magnify the differences among the bulk electric fields. The electric field in this region mainly causes the $\Omega$-pattern of the electron heating dynamics. This drift electric field becomes larger as the O$_2$ concentration increases, since the electron density gets lower, inducing an attenuation of the electron conductivity. It is known that the strong bulk electric field in atmospheric pressure discharges is caused by the very high collisionality of electrons, mainly due to electron elastic collisions, since cross sections of electron elastic collisions are much larger than those of electron inelastic collisions. In our study cases, the predominant contribution is electron helium elastic collisions due to the high flow of the helium gas. The increasing O$_2$ concentration from 0.05\% to 0.5\% cannot lead to a significant increase of the total collision frequency because O$_2$ is still a very small admixture and the cross section of electron oxygen elastic collisions is comparable with the one of electron helium elastic collisions. Attentively, here we only point out that the enhancement of the bulk electric field is not mainly caused by the elastic collisions, but by the increased electronegativity induced by adding more O$_2$. The right plot of figure \ref{fig8} shows the electric field perpendicular to the electrodes at t$_2$ = 19.2 ns corresponding to the moment of the maximal excitation rate in the top sheath (marked in the first row of figure \ref{fig3}). Again, only the region between 0.6 mm and 0.9 mm is shown to magnify the differences. The electric field in the top sheath becomes smaller as the O$_2$ concentration increases, which weakens the Penning-pattern of the electron heating dynamics there. Interestingly, the low electric field in the sheath leads to a self-amplification mechanism of this electron heating mode transition. Since the voltage is fixed across the discharge gap, the enhanced electric field in the bulk corresponds to a larger potential drop there. As a result, the voltage drop across the sheath is reduced, leading to the decreased electric field there.

\begin{figure}[t]
    \centering
    \includegraphics[width=0.6\textwidth]{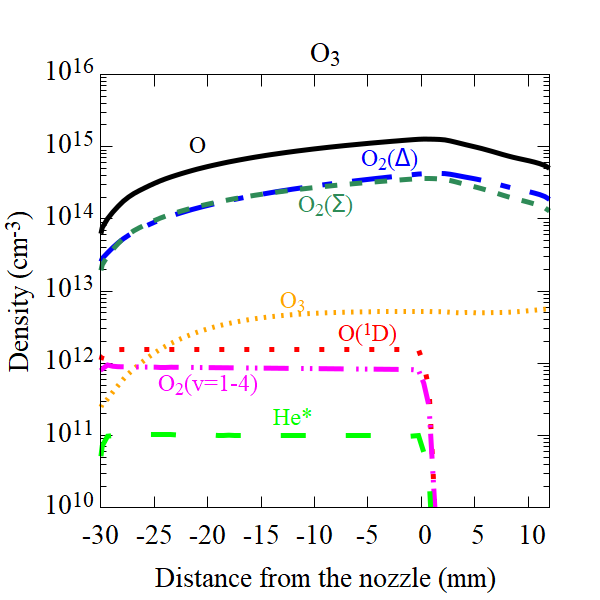}
    \caption{Computationally obtained distributions of neutral species densities at the center of the electrode gap along the direction of the gas flow (highlighted by the blue dotted line in figure \ref{fig1}). The gas nozzle is located at 0. The discharge domain is from -30 mm to 0. The effluent is at the position larger than 0. The voltage amplitude is 500 V and the O$_2$ concentration is 0.05\%.}
    \label{fig9}
\end{figure}

So far, we have discussed the electron heating mode transition dependence on the voltage amplitude and the O$_2$ concentration. We have also analyzed the charged species density variations induced by the control parameters and the effect of the electronegativity on the electron heating dynamics. Another important aspect for practical applications is the production of the desired reactive neutral species, such as atomic oxygen and ozone. We firstly focus on the computational distributions of neutral species densities along the direction of the gas flow at the fixed voltage amplitude of 500 V and the O$_2$ concentration of 0.05\%, as shown in figure \ref{fig9}. The densities are taken at the center of the electrode gap (highlighted by the blue dotted line in figure \ref{fig1}). Regarding the coordinate system, the gas nozzle is located at 0, indicating that the coordinate is negative inside the jet (discharge domain), while the coordinate of the effluent is positive. The He*, O($^1$D) and O$_2$(v=1-4) densities are flat distributed inside the jet, and rapidly decrease to be negligible after the nozzle. The O, O$_2$(a$^{1}\Delta_{g}$) and O$_2$(b$^{1}\Sigma_{g}^{+}$) densities increase in the jet and reach their highest values in a short distance from the nozzle outside the jet, and decrease slightly in the effluent. The O$_3$ density is continuously increasing until the end of the effluent propagation. Similar results have been reported by Hemke \textit{et al.} \cite{hemkeSpatiallyResolvedSimulation2011b}. Such neutral species distributions along the gas flow result from the major generation and destruction reaction rates, together with the effect of the gas flow. According to the sensitivity analyses based on chemical reactions in He/O$_2$ plasma jets in \cite{waskoenigAtomicOxygenFormation2010b,turnerUncertaintySensitivityAnalysis2015}, He*, O($^1$D) and O$_2$(v=1-4) are generated via electron impact excitation and quenched by He or O$_2$. Both processes are fast so that their density distributions are uniform in the jet. No source is provided outside the jet, so that they vanish rapidly. O, O$_2$(a$^{1}\Delta_{g}$) and O$_2$(b$^{1}\Sigma_{g}^{+}$) are also generated by electron impact molecular oxygen reactions at fast rates, but vanish slowly by reactions between neutrals. In such cases, the gas flow affects the transport of those species to show non-uniform distributions in the jet. Due to their relatively slow destruction rates, the species generated in the jet can be carried by the gas flow to the effluent. The decrease in the effluent is because of the lack of sources, but is not significant. O$_3$ is generated and destructed by neutrals slowly. The main generation reaction is

\begin{equation}
\mathrm O + \mathrm O_2 + \mathrm{He} \rightarrow \mathrm O_{3} + \mathrm{He},
\label{eq4}
\end{equation}
which provides a source in the effluent.

\begin{figure}[t]
    \centering
    \includegraphics[width=1.0\textwidth]{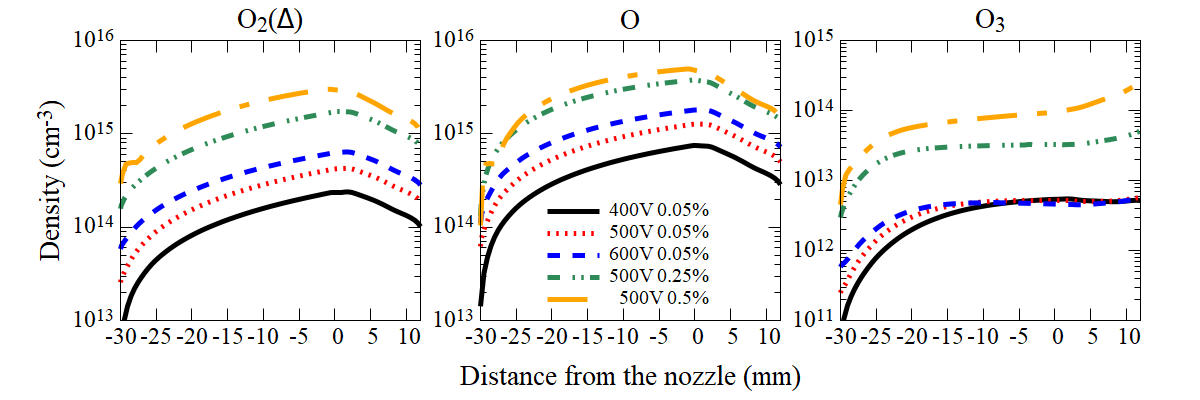}
    \caption{Computationally obtained distributions of reactive oxygen species densities at the center of the electrode gap along the direction of the gas flow (highlighted by the blue dotted line in figure \ref{fig1}). The gas nozzle is located at 0. The discharge domain is from -30 mm to 0. The effluent is located at the position larger than 0.}
    \label{fig10}
\end{figure}

Figure \ref{fig10} shows three reactive oxygen species (O$_2$(a$^{1}\Delta_{g}$), O and O$_3$) density distributions along the gas flow as a function of the voltage amplitude and the O$_2$ concentration. Densities are taken at the center of the electrode gap (highlighted by the blue dotted line in figure \ref{fig1}). Those species densities can be enhanced by either increasing the voltage amplitude or the O$_2$ concentration within the range discussed in this work. But increasing the O$_2$ admixtures is more effective.  

\section{\label{section4}Conclusions}
In this work, the spatio-temporal electron heating dynamics in a He/O$_2$ RF capacitive atmospheric pressure micro plasma jet is investigated by 2-dimensional fluid dynamics simulations and PROES measurements. Two electron heating modes (the $\Omega$-mode and the Penning-mode) are found and the electron heating dynamics can be changed by varying the voltage amplitude as well as the O$_2$ concentration. Based on the analyses of the construction and destruction mechanisms of the negative ions, it is pointed out that the generation of the negative ions is largely dependent on the O$_2$ concentration. The increased negative ion density induced by adding more O$_2$ enhances the drift bulk electric field, which contributes to the electron heating mode transition. The densities of different neutral species are found to show different distributions along the direction of the gas flow inside the jet and in the effluent due to the species relevant chemical reaction rates as well as the effect of the gas flow.

We notice that the electron heating mode transition can be induced within a lower range of the voltage amplitude in experiments (the second rows of figure \ref{fig2} and \ref{fig3}). However, our simulated results clearly show that the fluid model is capable of investigating RF micro atmospheric pressure discharges qualitatively and providing guidance for practical applications. To achieve quantitative agreement of the results between the electrodes with experimental measurements, a kinetic treatment of electrons is needed. Complex chemistry between heavy species can hardly be treated by Particle-in-Cell/Monte Carlo Collision (PIC/MCC) simulations to the best of our knowledge. A real hybrid model would be a more effective solution for simulating sophisticated discharges with multiple species more accurately. This has been already shown by the authors elsewhere based on a coupling between a 1d3v (one dimension in displacement, three dimensions in velocity) PIC/MCC model and a 2D simplified fluid model\cite{liuMicroAtmosphericPressure2020}. However, in order to consider the lateral gas flow, a sophisticated hybrid model including 2d3v PIC/MCC algorithm and 2D fluid model would be ultimately needed. Moreover, a large number of cells are required to resolve the plasma dynamics precisely in both directions. The implementation and application of such model is a future goal. In the meantime, one has to rely either on a hybrid 1d3v PIC/MCC model that allows for a correct kinetic description of electrons but omits the actual gas transport, or on the fluid model which considers the transport of all species, but treats electrons not fully kinetically.

\section*{Acknowledgments}
This work is supported by the German Research Foundation in the frame of SFB 1316 (project A4 and project A5) and MU2332/11-1. We gratefully thank Prof. Mark Kushner for his support and guidance of using \textit{nonPDPSIM}. We also thank Dr. Zoltán Donkó for fruitful discussions.

\section*{ORCID IDs}
\noindent T. Mussenbrock:  \href{http://orcid.org/0000-0001-6445-4990}{http://orcid.org/0000-0001-6445-4990}

\noindent I. Korolov: \href{https://orcid.org/0000-0003-2384-1243}{https://orcid.org/0000-0003-2384-1243}

\noindent J. Schulze: \href{https://orcid.org/0000-0001-7929-5734}{https://orcid.org/0000-0001-7929-5734}

\noindent Y. Liu: \href{https://orcid.org/0000-0002-2680-1338}{https://orcid.org/0000-0002-2680-1338}

\section*{References}

\bibliographystyle{ieeetr}
\bibliography{introduction.bib,costjet.bib,numericalmodel.bib,VWT.bib,RFplasmajet.bib,emissionfilter.bib,PIC-MCC.bib,dataO2.bib}

\begin{thebibliography}{10}

\bibitem{laroussiLowTemperaturePlasmaBased2005a}
M.~Laroussi, ``Low {{Temperature Plasma}}-{{Based Sterilization}}: {{Overview}}
  and {{State}}-of-the-{{Art}},'' {\em Plasma Processes and Polymers}, vol.~2,
  no.~5, pp.~391--400, 2005.

\bibitem{beckerMicroplasmasApplications2006a}
K.~H. Becker, K.~H. Schoenbach, and J.~G. Eden, ``Microplasmas and
  applications,'' {\em Journal of Physics D: Applied Physics}, vol.~39,
  pp.~R55--R70, Jan. 2006.

\bibitem{adamovich2017PlasmaRoadmap2017a}
I.~Adamovich, S.~D. Baalrud, A.~Bogaerts, P.~J. Bruggeman, M.~Cappelli,
  V.~Colombo, U.~Czarnetzki, U.~Ebert, J.~G. Eden, P.~Favia, D.~B. Graves,
  S.~Hamaguchi, G.~Hieftje, M.~Hori, I.~D. Kaganovich, U.~Kortshagen, M.~J.
  Kushner, N.~J. Mason, S.~Mazouffre, S.~M. Thagard, H.-R. Metelmann,
  A.~Mizuno, E.~Moreau, A.~B. Murphy, B.~A. Niemira, G.~S. Oehrlein, Z.~L.
  Petrovic, L.~C. Pitchford, Y.-K. Pu, S.~Rauf, O.~Sakai, S.~Samukawa,
  S.~Starikovskaia, J.~Tennyson, K.~Terashima, M.~M. Turner, M.~C.~M. van~de
  Sanden, and A.~Vardelle, ``The 2017 {{Plasma Roadmap}}: {{Low}} temperature
  plasma science and technology,'' {\em Journal of Physics D: Applied Physics},
  vol.~50, p.~323001, July 2017.

\bibitem{penkovReviewRecentApplications2015}
O.~V. Penkov, M.~Khadem, W.-S. Lim, and D.-E. Kim, ``A review of recent
  applications of atmospheric pressure plasma jets for materials processing,''
  {\em Journal of Coatings Technology and Research}, vol.~12, pp.~225--235,
  Mar. 2015.

\bibitem{babayanDepositionSiliconDioxide1998a}
S.~E. Babayan, J.~Y. Jeong, V.~J. Tu, J.~Park, G.~S. Selwyn, and R.~F. Hicks,
  ``Deposition of silicon dioxide films with an atmospheric-pressure plasma
  jet,'' {\em Plasma Sources Science and Technology}, vol.~7, pp.~286--288,
  Aug. 1998.

\bibitem{ichikiLocalizedUltrahighrateEtching2003a}
T.~Ichiki, R.~Taura, and Y.~Horiike, ``Localized and ultrahigh-rate etching of
  silicon wafers using atmospheric-pressure microplasma jets,'' {\em Journal of
  Applied Physics}, vol.~95, pp.~35--39, Dec. 2003.

\bibitem{kimCharacterizationAtmosphericPressure2009}
S.~J. Kim, T.~H. Chung, S.~H. Bae, and S.~H. Leem, ``Characterization of
  {{Atmospheric Pressure Microplasma Jet Source}} and its {{Application}} to
  {{Bacterial Inactivation}},'' {\em Plasma Processes and Polymers}, vol.~6,
  no.~10, pp.~676--685, 2009.

\bibitem{gravesLowTemperaturePlasma2014a}
D.~B. Graves, ``Low temperature plasma biomedicine: {{A}} tutorial review,''
  {\em Physics of Plasmas}, vol.~21, p.~080901, Aug. 2014.

\bibitem{weltmannPlasmaMedicineCurrent2016}
K.-D. Weltmann and T.~von Woedtke, ``Plasma medicine\textemdash current state
  of research and medical application,'' {\em Plasma Physics and Controlled
  Fusion}, vol.~59, p.~014031, Nov. 2016.

\bibitem{kongPlasmaMedicineIntroductory2009}
M.~G. Kong, G.~Kroesen, G.~Morfill, T.~Nosenko, T.~Shimizu, J.~van Dijk, and
  J.~L. Zimmermann, ``Plasma medicine: An introductory review,'' {\em New
  Journal of Physics}, vol.~11, p.~115012, Nov. 2009.

\bibitem{bekeschusPlasmaJetKINPen2016}
S.~Bekeschus, A.~Schmidt, K.-D. Weltmann, and T.~{von Woedtke}, ``The plasma
  jet {{kINPen}} \textendash{} {{A}} powerful tool for wound healing,'' {\em
  Clinical Plasma Medicine}, vol.~4, pp.~19--28, July 2016.

\bibitem{heilPossibilityMakingGeometrically2008a}
B.~G. Heil, U.~Czarnetzki, R.~P. Brinkmann, and T.~Mussenbrock, ``On the
  possibility of making a geometrically symmetric {{RF}}-{{CCP}} discharge
  electrically asymmetric,'' {\em Journal of Physics D: Applied Physics},
  vol.~41, p.~165202, July 2008.

\bibitem{schulzeElectricalAsymmetryEffect2011a}
J.~Schulze, E.~Sch{\"u}ngel, Z.~Donk{\'o}, and U.~Czarnetzki, ``The electrical
  asymmetry effect in multi-frequency capacitively coupled radio frequency
  discharges,'' {\em Plasma Sources Science and Technology}, vol.~20,
  p.~015017, Jan. 2011.

\bibitem{gibsonDisruptingSpatiotemporalSymmetry2019}
A.~R. Gibson, Z.~Donk{\'o}, L.~Alelyani, L.~Bischoff, G.~H{\"u}bner, J.~Bredin,
  S.~Doyle, I.~Korolov, K.~Niemi, T.~Mussenbrock, P.~Hartmann, J.~P. Dedrick,
  J.~Schulze, T.~Gans, and D.~O'Connell, ``Disrupting the spatio-temporal
  symmetry of the electron dynamics in atmospheric pressure plasmas by voltage
  waveform tailoring,'' {\em Plasma Sources Science and Technology}, vol.~28,
  p.~01LT01, Jan. 2019.

\bibitem{bischoffExperimentalComputationalInvestigations2018b}
L.~Bischoff, G.~H{\"u}bner, I.~Korolov, Z.~Donk{\'o}, P.~Hartmann, T.~Gans,
  J.~Held, V.~S.-v. der Gathen, Y.~Liu, T.~Mussenbrock, and J.~Schulze,
  ``Experimental and computational investigations of electron dynamics in micro
  atmospheric pressure radio-frequency plasma jets operated in {{He}}/{{N$_2$}}
  mixtures,'' {\em Plasma Sources Science and Technology}, vol.~27, p.~125009,
  Dec. 2018.

\bibitem{korolovControlElectronDynamics2019}
I.~Korolov, Z.~Donk{\'o}, G.~H{\"u}bner, L.~Bischoff, P.~Hartmann, T.~Gans,
  Y.~Liu, T.~Mussenbrock, and J.~Schulze, ``Control of electron dynamics,
  radical and metastable species generation in atmospheric pressure {{RF}}
  plasma jets by {{Voltage Waveform Tailoring}},'' {\em Plasma Sources Science
  and Technology}, vol.~28, p.~094001, Sept. 2019.

\bibitem{korolovHeliumMetastableSpecies2020}
I.~Korolov, M.~Leimk{\"u}hler, M.~B{\"o}ke, Z.~Donk{\'o}, V.~S.-v. der Gathen,
  L.~Bischoff, G.~H{\"u}bner, P.~Hartmann, T.~Gans, Y.~Liu, T.~Mussenbrock, and
  J.~Schulze, ``Helium metastable species generation in atmospheric pressure
  {{RF}} plasma jets driven by tailored voltage waveforms in mixtures of {{He}}
  and {{N$_2$}},'' {\em Journal of Physics D: Applied Physics}, vol.~53,
  p.~185201, Feb. 2020.

\bibitem{liuMicroAtmosphericPressure2020}
Y.~Liu, I.~Korolov, J.~Trieschmann, D.~Steuer, V.~S.-v. der Gathen, M.~Boeke,
  L.~Bischoff, G.~H{\"u}bner, J.~Schulze, and T.~Mussenbrock, ``Micro
  atmospheric pressure plasma jets excited in {{He}}/{{O$_2$}} by voltage
  waveform tailoring: {{A}} study based on a numerical hybrid model and
  experiments,'' {\em Plasma Sources Science and Technology}, 2020.

\bibitem{izaElectronKineticsRadioFrequency2007a}
F.~Iza, J.~K. Lee, and M.~G. Kong, ``Electron {{Kinetics}} in
  {{Radio}}-{{Frequency Atmospheric}}-{{Pressure Microplasmas}},'' {\em
  Physical Review Letters}, vol.~99, p.~075004, Aug. 2007.

\bibitem{niemiRoleHeliumMetastable2011}
K.~Niemi, J.~Waskoenig, N.~Sadeghi, T.~Gans, and D.~O'Connell, ``The role of
  helium metastable states in radio-frequency driven helium\textendash oxygen
  atmospheric pressure plasma jets: Measurement and numerical simulation,''
  {\em Plasma Sources Science and Technology}, vol.~20, p.~055005, Aug. 2011.

\bibitem{niermannSpatialDynamicsHelium2011b}
B.~Niermann, T.~Hemke, N.~Y. Babaeva, M.~B{\"o}ke, M.~J. Kushner,
  T.~Mussenbrock, and J.~Winter, ``Spatial dynamics of helium metastables in
  sheath or bulk dominated rf micro-plasma jets,'' {\em Journal of Physics D:
  Applied Physics}, vol.~44, p.~485204, Nov. 2011.

\bibitem{dunnbierStabilityExcitationDynamics2015}
M.~D{\"u}nnbier, M.~M. Becker, S.~Iseni, R.~Bansemer, D.~Loffhagen, S.~Reuter,
  and K.-D. Weltmann, ``Stability and excitation dynamics of an argon
  micro-scaled atmospheric pressure plasma jet,'' {\em Plasma Sources Science
  and Technology}, vol.~24, p.~065018, Nov. 2015.

\bibitem{chirokovNumericalExperimentalInvestigation2009}
A.~Chirokov, S.~N. Khot, S.~P. Gangoli, A.~Fridman, P.~Henderson, A.~F. Gutsol,
  and A.~Dolgopolsky, ``Numerical and experimental investigation of the
  stability of radio-frequency ({{RF}}) discharges at atmospheric pressure,''
  {\em Plasma Sources Science and Technology}, vol.~18, p.~025025, Mar. 2009.

\bibitem{kawamuraParticleincellGlobalSimulations2014a}
E.~Kawamura, M.~A. Lieberman, A.~J. Lichtenberg, P.~Chabert, and C.~Lazzaroni,
  ``Particle-in-cell and global simulations of $\alpha$ to $\gamma$ transition
  in atmospheric pressure {{Penning}}-dominated capacitive discharges,'' {\em
  Plasma Sources Science and Technology}, vol.~23, p.~035014, May 2014.

\bibitem{hemkeIonizationBulkHeating2012a}
T.~Hemke, D.~Eremin, T.~Mussenbrock, A.~Derzsi, Z.~Donk{\'o}, K.~Dittmann,
  J.~Meichsner, and J.~Schulze, ``Ionization by bulk heating of electrons in
  capacitive radio frequency atmospheric pressure microplasmas,'' {\em Plasma
  Sources Science and Technology}, vol.~22, p.~015012, Dec. 2012.

\bibitem{martensDominantRoleImpurities2008}
T.~Martens, A.~Bogaerts, W.~J.~M. Brok, and J.~V. Dijk, ``The dominant role of
  impurities in the composition of high pressure noble gas plasmas,'' {\em
  Applied Physics Letters}, vol.~92, p.~041504, Jan. 2008.

\bibitem{raduFrequencyVoltageDependence2003}
I.~Radu, R.~Bartnikas, and M.~R. Wertheimer, ``Frequency and voltage dependence
  of glow and pseudoglow discharges in helium under atmospheric pressure,''
  {\em IEEE Transactions on Plasma Science}, vol.~31, pp.~1363--1378, Dec.
  2003.

\bibitem{gathenSpatiallyResolvedDiagnostics2008}
V.~S.-v. der Gathen, L.~Schaper, N.~Knake, S.~Reuter, K.~Niemi, T.~Gans, and
  J.~Winter, ``Spatially resolved diagnostics on a microscale atmospheric
  pressure plasma jet,'' {\em Journal of Physics D: Applied Physics}, vol.~41,
  p.~194004, Sept. 2008.

\bibitem{benediktPhaseResolvedOptical2010}
J.~Benedikt, S.~Hofmann, N.~Knake, H.~B{\"o}ttner, R.~Reuter, A.~{von Keudell},
  and V.~{Schulz-von der Gathen}, ``Phase resolved optical emission
  spectroscopy of coaxial microplasma jet operated with {{He}} and {{Ar}},''
  {\em The European Physical Journal D}, vol.~60, pp.~539--546, Dec. 2010.

\bibitem{reuterDetectionOzoneMHz2012}
S.~Reuter, J.~Winter, S.~Iseni, S.~Peters, A.~{Schmidt-Bleker},
  M.~D{\"u}nnbier, J.~Sch{\"a}fer, R.~Foest, and K.-D. Weltmann, ``Detection of
  ozone in a {{MHz}} argon plasma bullet jet,'' {\em Plasma Sources Science and
  Technology}, vol.~21, p.~034015, May 2012.

\bibitem{schaperElectronDynamicsRadioFrequencyDriven2011}
L.~Schaper, J.~Waskoenig, M.~G. Kong, V.~S.-v. der Gathen, and T.~Gans,
  ``Electron {{Dynamics}} in a {{Radio}}-{{Frequency}}-{{Driven
  Microatmospheric Pressure Plasma Jet}},'' {\em IEEE Transactions on Plasma
  Science}, vol.~39, pp.~2370--2371, Nov. 2011.

\bibitem{grebEnergyResolvedActinometry2014}
A.~Greb, K.~Niemi, D.~O'Connell, and T.~Gans, ``Energy resolved actinometry for
  simultaneous measurement of atomic oxygen densities and local mean electron
  energies in radio-frequency driven plasmas,'' {\em Applied Physics Letters},
  vol.~105, p.~234105, Dec. 2014.

\bibitem{waskoenigAtomicOxygenFormation2010b}
J.~Waskoenig, K.~Niemi, N.~Knake, L.~M. Graham, S.~Reuter, V.~S.-v. der Gathen,
  and T.~Gans, ``Atomic oxygen formation in a radio-frequency driven
  micro-atmospheric pressure plasma jet,'' {\em Plasma Sources Science and
  Technology}, vol.~19, p.~045018, June 2010.

\bibitem{belenguerTransitionDifferentRegimes1990}
P.~Belenguer and J.~P. Boeuf, ``Transition between different regimes of rf glow
  discharges,'' {\em Physical Review A}, vol.~41, pp.~4447--4459, Apr. 1990.

\bibitem{brinkmannElectricFieldCapacitively2015}
R.~P. Brinkmann, ``The electric field in capacitively coupled {{RF}}
  discharges: A smooth step model that includes thermal and dynamic effects,''
  {\em Plasma Sources Science and Technology}, vol.~24, p.~064002, Oct. 2015.

\bibitem{brinkmannElectronHeatingCapacitively2015}
R.~P. Brinkmann, ``Electron heating in capacitively coupled {{RF}} plasmas: A
  unified scenario,'' {\em Plasma Sources Science and Technology}, vol.~25,
  p.~014001, Dec. 2015.

\bibitem{schulzeSpatiotemporalAnalysisElectron2018}
J.~Schulze, Z.~Donk{\'o}, T.~Lafleur, S.~Wilczek, and R.~P. Brinkmann,
  ``Spatio-temporal analysis of the electron power absorption in
  electropositive capacitive {{RF}} plasmas based on moments of the
  {{Boltzmann}} equation,'' {\em Plasma Sources Science and Technology},
  vol.~27, p.~055010, May 2018.

\bibitem{schulzeIonizationDriftAmbipolar2011b}
J.~Schulze, A.~Derzsi, K.~Dittmann, T.~Hemke, J.~Meichsner, and Z.~Donk{\'o},
  ``Ionization by {{Drift}} and {{Ambipolar Electric Fields}} in
  {{Electronegative Capacitive Radio Frequency Plasmas}},'' {\em Physical
  Review Letters}, vol.~107, p.~275001, Dec. 2011.

\bibitem{hemkeSpatiallyResolvedSimulation2011b}
T.~Hemke, A.~Wollny, M.~Gebhardt, R.~P. Brinkmann, and T.~Mussenbrock,
  ``Spatially resolved simulation of a radio-frequency driven micro-atmospheric
  pressure plasma jet and its effluent,'' {\em Journal of Physics D: Applied
  Physics}, vol.~44, p.~285206, June 2011.

\bibitem{turnerUncertaintySensitivityAnalysis2015}
M.~M. Turner, ``Uncertainty and sensitivity analysis in complex plasma
  chemistry models,'' {\em Plasma Sources Science and Technology}, vol.~25,
  p.~015003, Dec. 2015.

\bibitem{turnerUncertaintyErrorComplex2015a}
M.~M. Turner, ``Uncertainty and error in complex plasma chemistry models,''
  {\em Plasma Sources Science and Technology}, vol.~24, p.~035027, June 2015.

\bibitem{wijaikhumAbsoluteOzoneDensities2017a}
A.~Wijaikhum, D.~Schr{\"o}der, S.~Schr{\"o}ter, A.~R. Gibson, K.~Niemi,
  J.~Friderich, A.~Greb, V.~S.-v. der Gathen, D.~O'Connell, and T.~Gans,
  ``Absolute ozone densities in a radio-frequency driven atmospheric pressure
  plasma using two-beam {{UV}}-{{LED}} absorption spectroscopy and numerical
  simulations,'' {\em Plasma Sources Science and Technology}, vol.~26,
  p.~115004, Oct. 2017.

\bibitem{babaevaProductionO2D12006a}
N.~Y. Babaeva, R.~A. Arakoni, and M.~J. Kushner, ``Production of
  {{O$_2$}}({{$^1\Delta$}}) in flowing plasmas using spiker-sustainer
  excitation,'' {\em Journal of Applied Physics}, vol.~99, p.~113306, June
  2006.

\bibitem{babaevaO2D1Production2007b}
N.~Y. Babaeva, R.~Arakoni, and M.~J. Kushner, ``O$_2$({{$^1\Delta$}})
  production in high pressure flowing {{He}}/{{O$_2$}} plasmas: {{Scaling}} and
  quenching,'' {\em Journal of Applied Physics}, vol.~101, p.~123306, June
  2007.

\bibitem{lietzElectrodeConfigurationsAtmospheric2018}
A.~M. Lietz and M.~J. Kushner, ``Electrode configurations in atmospheric
  pressure plasma jets: Production of reactive species,'' {\em Plasma Sources
  Science and Technology}, vol.~27, p.~105020, Oct. 2018.

\bibitem{norbergFormationReactiveOxygen2015}
S.~A. Norberg, E.~Johnsen, and M.~J. Kushner, ``Formation of reactive oxygen
  and nitrogen species by repetitive negatively pulsed helium atmospheric
  pressure plasma jets propagating into humid air,'' {\em Plasma Sources
  Science and Technology}, vol.~24, p.~035026, June 2015.

\bibitem{norbergAtmosphericPressurePlasma2018}
S.~A. Norberg, G.~M. Parsey, A.~M. Lietz, E.~Johnsen, and M.~J. Kushner,
  ``Atmospheric pressure plasma jets onto a reactive water layer over tissue:
  Pulse repetition rate as a control mechanism,'' {\em Journal of Physics D:
  Applied Physics}, vol.~52, p.~015201, Oct. 2018.

\bibitem{kushnerModelingMicrodischargeDevices2004a}
M.~J. Kushner, ``Modeling of microdischarge devices: {{Pyramidal}}
  structures,'' {\em Journal of Applied Physics}, vol.~95, pp.~846--859, Jan.
  2004.

\bibitem{kushnerModellingMicrodischargeDevices2005a}
M.~J. Kushner, ``Modelling of microdischarge devices: Plasma and gas
  dynamics,'' {\em Journal of Physics D: Applied Physics}, vol.~38,
  pp.~1633--1643, May 2005.

\bibitem{goldaConceptsCharacteristicsCOST2016}
J.~Golda, J.~Held, B.~Redeker, M.~Konkowski, P.~Beijer, A.~Sobota, G.~Kroesen,
  N.~S.~J. Braithwaite, S.~Reuter, M.~M. Turner, T.~Gans, D.~O'Connell, and
  V.~S.-v. der Gathen, ``Concepts and characteristics of the `{{COST Reference
  Microplasma Jet}}','' {\em Journal of Physics D: Applied Physics}, vol.~49,
  p.~084003, Jan. 2016.

\bibitem{knakeAbsoluteAtomicOxygen2008a}
N.~Knake, K.~Niemi, S.~Reuter, V.~{Schulz-von der Gathen}, and J.~Winter,
  ``Absolute atomic oxygen density profiles in the discharge core of a
  microscale atmospheric pressure plasma jet,'' {\em Applied Physics Letters},
  vol.~93, p.~131503, Sept. 2008.

\bibitem{CrossSectionsExtracted}
``Cross sections extracted from {{PROGRAM MAGBOLTZ}}, {{VERSION}} 7.1 {{JUNE}}
  2004.'' www.lxcat.net/Biagi-v7.1.

\bibitem{CrossSectionsExtractedc}
``Cross sections extracted from {{MAGBOLTZ}}, {{Biagi S F}}, version 8.9.''
  www.lxcat.net/Biagi.

\bibitem{gudmundssonBenchmarkStudyCapacitively2013}
J.~T. Gudmundsson, E.~Kawamura, and M.~A. Lieberman, ``A benchmark study of a
  capacitively coupled oxygen discharge of the oopd1 particle-in-cell {{Monte
  Carlo}} code,'' {\em Plasma Sources Science and Technology}, vol.~22,
  p.~035011, May 2013.

\bibitem{staffordO2D1Production2004b}
D.~S. Stafford and M.~J. Kushner, ``O$_2$({{$^1\Delta$}}) production in
  {{He}}/{{O$_2$}} mixtures in flowing low pressure plasmas,'' {\em Journal of
  Applied Physics}, vol.~96, pp.~2451--2465, Sept. 2004.

\bibitem{raizerGasDischargePhysics1991}
Y.~P. Raizer, {\em Gas {{Discharge Physics}}}.
\newblock {Berlin Heidelberg}: {Springer-Verlag}, 1991.

\end{thebibliography}

\end{document}